\documentclass[journal=jctcce,manuscript=article]{achemso}

\usepackage{amsmath}
\usepackage{bm}
\usepackage[dvipsnames]{xcolor}
\usepackage[version=3]{mhchem} 
\usepackage{multirow}
\usepackage{hyperref}
\usepackage{pdflscape}
\usepackage{physics}
\usepackage{siunitx}
\usepackage{upgreek}
\usepackage{xparse}
\usepackage{xspace}

\newcommand*{\XPi}{\mathrm{X}\,^2\Pi}
\newcommand*{\XSigma}{\mathrm{X}\,^2\Sigma^+}

\newcommand*{\eg}{\latin{e.g.}\xspace}
\newcommand*{\etal}{\latin{et al.}\xspace}
\newcommand*{\etc}{\latin{etc.}\xspace}
\newcommand*{\ie}{\latin{i.e.}\xspace}

\newcommand*{\duo}{\textsc{Duo}\xspace}

\newcommand*{\shrodinger}{{Schr\"odinger}\xspace}
\newcommand*{\state}{\mathrm{state}}
\newcommand*{\mat}[1]{\bm{#1}}
\newcommand*{\Ldoubling}{{$\varLambda$\,-\,doubling}\xspace}
\newcommand*{\abinitio}{\textit{ab~initio}\xspace}
\newcommand*{\reducedmel}[3]{\left\langle{#1}\norm{#2}{#3}\right\rangle\xspace}
\newcommand*{\tensorT}{\mathrm{T}}
\newcommand*{\rotateD}{\mathcal{D}}
\renewcommand{\ell}{l}
\newcommand*{\rotateDelement}[3]{\rotateD_{#1}^{(#2)}(#3)\xspace}
\newcommand*{\WignerThreej}[6]{
    \begin{pmatrix}
        #1 & #2 & #3 \\
        #4 & #5 & #6
    \end{pmatrix}
}
\newcommand*{\WignerSixj}[6]{
    \begin{Bmatrix}
        #1 & #2 & #3 \\
        #4 & #5 & #6
    \end{Bmatrix}
}
\newcommand*{\sphTensorOp}[3]{\tensorT_{#1}^{#2}(#3)\xspace}

\newcommand*{\opH}{\mathcal{H}}
\newcommand*{\Hhfs}{\opH_\mathrm{hfs}}
\newcommand*{\HFC}{\opH_\mathrm{FC}}
\newcommand*{\HIL}{\opH_\mathrm{IL}}

\newcommand*{\HEQ}{\opH_\mathrm{EQ}}
\newcommand*{\HIJ}{\opH_\mathrm{IJ}}
\newcommand*{\Hdip}{\opH_\mathrm{dip}}

\newcommand*{\Hzero}{\opH^{(0)}}

\newcommand*{\red}[1]{{\color{red} #1}}

\newcommand*{\bms}{\bm{s}}
\newcommand*{\bmS}{\bm{S}}
\newcommand*{\bmJ}{\bm{J}}
\newcommand*{\bmL}{\bm{L}}
\newcommand*{\bml}{\bm{\ell}}
\newcommand*{\bmI}{\bm{I}}
\newcommand*{\bmR}{\bm{R}}
\newcommand*{\bmQ}{\bm{Q}}
\newcommand*{\bmC}{\bm{C}}
\newcommand*{\bmE}{\bm{E}}
\newcommand*{\bmr}{\bm{r}}

\newcommand*{\bmmu}{\bm{\mu}}
\newcommand*{\bmmutrans}{\bm{\mu}}

\newcommand*{\bmF}{\bm{F}}

\newcommand*{\bmomega}{\bm{\omega}}

\newcommand*{\DiracDelta}[1]{\updelta({#1})}
\newcommand*{\kronecker}[1]{\updelta_{#1}}
\newcommand*{\Identity}[2]{\left(\sum_{#1}\dyad{#2}\right)}

\newcommand*{\bF}{b_\mathrm{F}}

\newcommand*{\parity}{\tau}
\newcommand*{\constantN}{ g_N\, \mu_B\, \mu_N\, \frac{\mu_0}{4 \uppi}}

\author{Qianwei Qu}
\author{Sergei N. Yurchenko}
\author{Jonathan Tennyson}
\email{j.tennyson@ucl.ac.uk}
\affiliation{Department of Physics and Astronomy,
University College London,
WC1E 6BT London, UK}

\title[Diatomic hyperfine spectra]
{A method for the variational calculation of hyperfine-resolved rovibronic spectra  
of diatomic molecules}

\abbreviations{IR,NMR,UV}

\begin{document}

\begin{tocentry}
\centering
\includegraphics{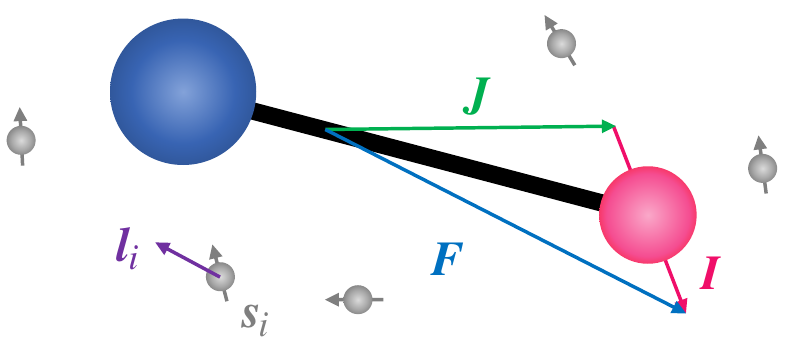}

\end{tocentry}

{
We thank \href{https://www.bristol.ac.uk/news/2021/october/colin-western.html}{Colin Western (1957-2021)}, who created and maintained 
PGOPHER, for
the immense help he provided to this and others studies of ours. We
wish to dedicate this paper to his memory.}

\red{The manuscript has been accepted by
\emph{Journal of Chemical Theory and Computation}.
}

\begin{abstract}
An algorithm for the calculation of 
hyperfine structure and spectra of diatomic molecules based on the
variational nuclear motion is presented.
Hyperfine coupling terms considered are
Fermi-contact, 
nuclear spin-electron spin dipole-dipole, 
nuclear spin-orbit, 
nuclear spin-rotation
and nuclear electric quadrupole 
interactions. 
Initial hyperfine-unresolved wavefunctions are obtained for
given set of potential energy curves and associated couplings
by variation solution of the nuclear-motion
Schr\"odinger equation.
Fully hyperfine-resolved parity-conserved
rovibronic Hamiltonian matrices
for a given final angular momentum, $\bm{F}$,
are constructed and then
 diagonalized to give 
hyperfine-resolved energies and wavefunctions.
Electric transition dipole moment curves can then be used to generate
a hyperfine-resolved line list by
applying rigorous selection rules.
The algorithm is implemented in \textsc{Duo},
which is a general program for calculating 
spectra of diatomic molecules.
This approach is tested for 
NO and MgH,
and the results are compared to experiment and shown to be consistent with
those given the well-used effective Hamiltonian code PGOPHER.
\end{abstract}

\section{Introduction}
The hyperfine structure of  molecules
lays the foundation for 
the studies of many important areas.
The most immediate application
is to reveal the properties of the molecules
\cite{20PuKoPa.hyperfine,21FaMexx.hpyerfine,12BrChCh.hyperfine}{}.
Other examples includes
laser cooling experiments \cite{13HuYeSt.hyperfine,15YeHuCo.hyperfine}{}, astronomical observations \cite{06HaBrHa.hyperfine}{}, 
and, of course, nuclear magnetic resonance which has many applications including ones in medicine.

In the absence of external fields, the rotational hyperfine structure 
results from interactions between
the electric and magnetic multipole moments
of the nuclei and their molecular environments 
\cite{71CoLuxx.hyperfine}{}.
Due to parity conservation inside the nuclei, 
only even electric and 
odd magnetic multipoles are non vanishing.
Although higher multipole effects are observed in some experiments,
the dominant contributions to the hyperfine structure
arise from magnetic dipole and
electric quadrupole interactions.

Frosch and Foley\cite{52FrFoxx.hyperfine} performed a pioneering theoretical study
of the magnetic interactions between nuclei and 
electron spins in diatomic molecules
based on the Dirac equation, see discussion by
Brown and Carrington\cite{03BrCaxx.method}{}.
\citeauthor{48BaToxx.hyperfine}  \cite{48BaToxx.hyperfine} provided the first extensive discussion of the electric quadrupole interactions.

The application of irreducible spherical tensor operators
facilitate the evaluation of 
effective hyperfine Hamiltonian matrix elements
\cite{66Freedx.hyperfine,71CoLuxx.hyperfine,78BrViLe.hyperfine,93Kato.methods,03BrCaxx.method}{},
although one must still pay attention to 
anomalous commutation relationships
when coupling angular momenta
\cite{76BrHoxx.method,51Vanvleck.linear}{}.
Standard practice is to use these matrix elements
to solve problems where hyperfine structure is important using 
effective Hamiltonians  which implicitly use a perturbation theory based representation
of the problem \cite{06HaBrHa.hyperfine,12BrChCh.hyperfine}{}.
The effective Hamiltonian of a fine or hyperfine problem
is usually constructed within a particular vibrational state and 
the rotational coupling terms are treated as perturbations.
The assumptions implicit in this approach are usually valid 
because the splitting of the (rotational) energy levels  due to hyperfine effects are generally small
compared to the separation between electronic or vibrational states.
However, this assumption can fail, such as  for example, for Rydberg
states of molecules\cite{04OsWuMe.hyperfine,20DeHoxx.NO}{}.
The B~$^2\Pi$ -- C~$^2\Pi$ avoided crossing structure in NO is another example of 
strong electronic state interaction.
The perturbative treatment of this vibronic coupling
is difficult: it  requires a lot of parameters \cite{82GaDrxx.NO}{},
and is not very accurate.
The interaction between different states leads to significant complications which are difficult
to model using the standard effective Hamiltonian approach.

In contrast 
our recent work on a spectroscopic model for the four lowest electronic states of NO \cite{jt821}  
proposed a compact solution for the problem based on the use of a variational method to treat the nuclear motion.
In our approach, which is bases on the use of potential energy curves and appropriate couplings, it was only necessary
to introduce one potential energy coupling curve between the
coupled B~$^2\Pi$ and C~$^2\Pi$ electronic states; this
 gave an accurate rovibronic line list for NO \cite{jt845}{}.
These calculations used  
a general program for the calculation of spectra of diatomic molecules,
\duo \cite{jt609}{}.

\duo is a variational nuclear motion program developed for the calculation of rovibronic
spectra of diatomic molecules as part of the ExoMol project\cite{jt626}{}. 
It provides explicit treatment of spin-orbit and other coupling terms
and can generate high-accuracy fine-structure diatomic line lists.
\duo has been used to generate many line lists including those
for \ce{AlO} \cite{jt598}{}, 
\ce{CaO} \cite{jt618}{},
\ce{VO} \cite{jt644}{}, 
\ce{TiO} \cite{jt760},
\ce{YO} \cite{jt774}{},
and \ce{SiO} \cite{jt847}{},
which are provided via the ExoMol database \cite{jt810}{}.
\duo was also recently employed to calculate temperature-dependent photodissociation cross sections and rates
\cite{jt840}{}. \duo has also been adapted to treat ultra-low energy collisions
as the inner region in an R-matrix formalism \cite{jt755}{}; hyperfine effects
are very important in such collisions.
Recently, 
a new module treating electric quadrupole transitions
has been added to \duo\cite{21SoYuYa}{},
which makes it capable of predicting
spectra for diatomic molecules with no electric dipole moment,
\eg \ce{O2} and \ce{N2}.
However, up until now
\duo does not  treated hyperfine effects.
In this context we note
that hyperfine coupling is particularly strong for VO \cite{95AdBaBe.VO, 08FiZixx.VO} meaning that
the current ExoMol VO line list, VOMYT \cite{jt644} which is not hyperfine resolved, is unsuitable for 
high resolution work such studies of exoplanets using high-resoluton Doppler-shift spectroscopy \cite{20MeGiNu.VO}.

Here we present a variational procedure for calculating hyperfine-resolved spectra of diatomic molecules.
The new algorithm we design is implemented as new modules in \duo.
In general,
the most challenging part of solving quantum mechanical problems using a
variational method is finding good variational basis sets.
We show below 
that \duo  gives appropriate basis sets thanks to its 
well-designed calculation hierarchy and algorithm.
Numerical tests  indicates that the algorithm proposed here
can achieve high accuracy for calculation of hyperfine structure.

\section{Overview}
\label{sec:overview}

\begin{figure}
    \centering
    \includegraphics{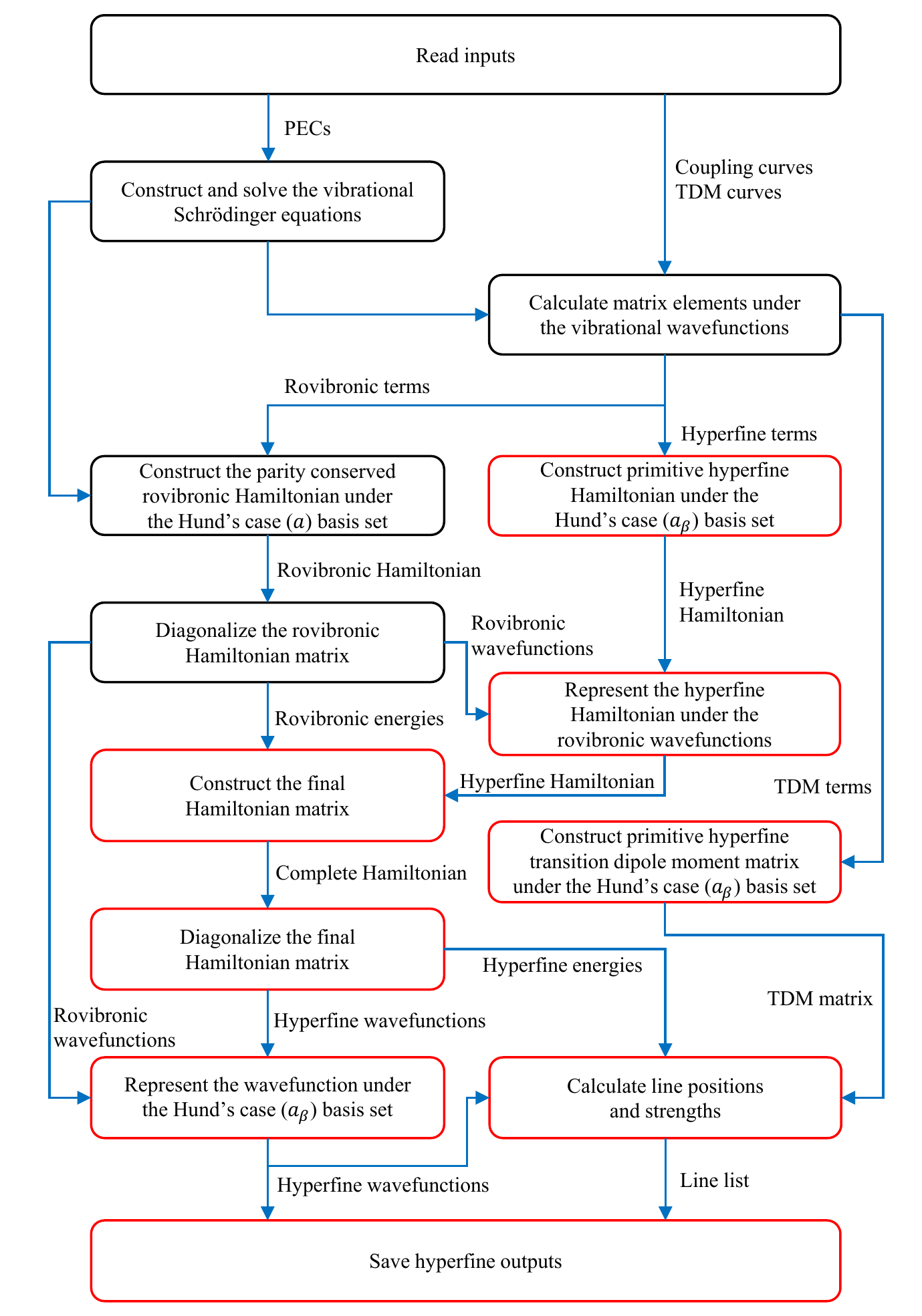}
    \caption{Flowchart showing the structure of a \duo hyperfine  calculation.
    Existing modules are given by black rectangles while new modules are denoted 
    by red rectangles.
    PEC is short for potential energy curve
    and TDM is short for transition dipole moment.}
    \label{fig:flowchart}
\end{figure}

In this section,
we outline our algorithm
so that the readers can easily
follow the details given in the following sections. Figure~\ref{fig:flowchart} gives
a graphical representation of the algorithm.

We write the Hamiltonian for the problem as 
\begin{equation}
    \opH =
        \Hzero +
        \Hhfs,
\end{equation}
where $\Hzero$ is the  rovibronic Hamiltonian 
which \duo originally used 
to give fine structure resolved solutions for diatomic molecules,
and $\Hhfs$ gives the nuclear hyperfine interaction terms 
introduced in this work.
We emphasize that 
although this structure is the standard one used in perturbation theory, here
we aim for a full variational solution of the whole Hamiltonian $\opH$.

\subsection{Rovibronic fine structure}
\label{sec:finestructure}
\duo has well-developed modules,
surrounded by black rectangles in Fig.\,\ref{fig:flowchart},
for the calculation of
rovibronic energies and wavefunctions.

The computational procedure used by \duo to obtain solutions for $\Hzero$
is divided in two steps. 
First, 
the rotationless \shrodinger equation is solved independently for each uncoupled potential energy curve, ${V}_\state(R)$,
to give vibrational energy levels, $E_{\state,v}$,
and wavefunctions, $\psi_{\state, v}$\,:
\begin{equation}
    -\frac{\hbar^2}{2\mu}\dv[2]{R} \psi_{\state, v}(R)+
    {V}_\state(R)
    \psi_{\state, v}(R)
    = E_{\state,v}\,\psi_{\state, v}(R)\,,
    \label{eq:vibeq}
\end{equation}
where $R$ is the internuclear distance,
$\mu$ is the reduced mass of the molecule,
`$\state$' and $v$ indicate
the electronic state and vibrational
quantum numbers.
\duo employs contracted vibrational basis sets given by $\psi_{\state, v} = \ket{\state, v}$
to define a finite-dimension space.

In the second step,
a rovibronic Hamiltonian matrix,
corresponding to $\Hzero$,
for each specific total angular momentum exclusive of nuclear spin, $J$,
and parity, $\parity$,
is constructed
using a Hund's case (a) basis set \cite{jt632}{}:
\begin{equation}
    \ket{\state,v,\varLambda,S,\varSigma, J,\varOmega}
    = \ket{\state, \varLambda, S, \varSigma}\ket{\state, v} 
        \ket{J,\varOmega, M_J},
    \label{eq:casea}
\end{equation}
which is decoupled into three parts:
(i) the electronic eigenfunction,
(ii) the vibrational eigenfunction of Eq.\,(\ref{eq:vibeq}),
and (iii) the rotational eigenfunction of a symmetric top.
The quantum numbers in Eq.\,(\ref{eq:casea}),
$\state$, $v$, $\varLambda$, $S$, $\varSigma$
$J$, $\varOmega$ and $M_J$,
correspond to 
the electronic state,
the vibrational eigenstate,
the projection of the electron orbital angular momentum $\bmL$ on the molecular axis,
the projection of the electron spin angular momentum $\bmS$ on the molecular axis,
the projection of $\bmJ$ on the molecular axis and 
the projection of $\bmJ$ on the space-fixed $Z$-axis,
respectively.
Note that,
\duo calculates
the spectra of diatomic molecules in field-free environments.
Thus,
we do not really use $M_J$ to construct the basis set,
as the left hand side of Eq.\,(\ref{eq:casea}) indicates.
All the angular momenta are quantized to the body-fixed axes.

When evaluating the matrix elements 
using the basis functions of Eq.\,(\ref{eq:casea}), 
the necessary coupling curves are
integrated over pairs
of vibrational basis functions:
\begin{equation}
    ^{\state,v}C^{\state',v'}
    =\mel**{\state, v}{^{\state}C^{\state'}(R)}{\state', v'},
\end{equation}
where $C(R)$ can be either a diagonal coupling curve 
for a particular electronic state or
an off-diagonal coupling curve between two states.
Supported couplings include
electron spin-orbit,
electron spin-spin,
electron spin-rotation
\etc
\cite{jt609,jt632}.

The basis functions of Eq.\,(\ref{eq:casea}) do not
have definite parities.
\duo uses linear combinations of them to define parity-conserved basis functions:
\begin{equation}
\begin{aligned}
    &+: \frac{1}{\sqrt{2}}\ket{\state,v,\varLambda,S,\varSigma, J,\varOmega} +
    \frac{1}{\sqrt{2}}(-1)^{s-\varLambda+S-\varSigma+J-\varOmega}
    \ket{\state,v,-\varLambda,S,-\varSigma, J,-\varOmega}, \\
    &-: \frac{1}{\sqrt{2}}\ket{\state,v,\varLambda,S,\varSigma, J,\varOmega}
    -\frac{1}{\sqrt{2}}(-1)^{s-\varLambda+S-\varSigma+J-\varOmega}\ket{\state,v,-\varLambda,S,-\varSigma, J,-\varOmega},
\end{aligned}
\label{eq:parityconserved}
\end{equation}
where $s=1$
for $\Sigma^-$ states and 
$s = 0$ for all other states.
Note that,
the parity is independent of $M_J$.
Each matrix of $\Hzero$ constructed using these basis functions 
can be diagonalized 
to give rovibronic energy levels and
wavefunctions of a definite $J$ and parity $\parity$.
Let $\ket{\phi^{\parity,J}_m}$ be
the $m$-th eigenfunction corresponding to
a given $J$ and parity $\parity$,
we have:
\begin{equation}
    \mel**{\phi^{\parity,J}_m}{\Hzero}{\phi^{\parity,J}_{m'}} 
    = \kronecker{m,m'}\,  E_{m}^{\tau,J},
\end{equation}
where $ E_{m}^{\tau,J}$ is the $m$-th eigenvalue.

Thanks to the use of complete angular basis sets and the variational method,
the final energies are independent
of the coupling scheme used. 
If enough vibrational basis 
(determined by the users' setup),
the choice of Hund's case (a) will 
give correct results even for cases
where other coupling schemes 
provide a better zeroth-order approximation.

\subsection{Nuclear hyperfine structure}
We program new \duo modules
to accomplish the functions denoted by
the red rectangles in Fig.\,\ref{fig:flowchart}
for nuclear hyperfine structure calculations.
We only consider heteronuclear
diatomic molecules 
with one nucleus possessing non-zero spin
in this paper. In this case,
nuclear spin, $\bmI$,
is coupled with $\bmJ$ to give 
total angular momentum, $\bmF$, 
\ie, 
\begin{equation}
    \bmF = \bmI + \bmJ.
\end{equation}
To evaluate the matrix elements of $\Hhfs$,
    we introduce the following primitive basis functions 
    \begin{equation}
        \ket{\state, v, \varLambda, S,\varSigma,J,\varOmega, I,F,M_F}
        = 
         \ket{\state,\varLambda,S,\varSigma}\ket{\state,v}\ket{J,\varOmega,M_J}
            \ket{J, I, F,M_F}
         \label{eq:rovibronic_basis:0},
    \end{equation}
 where   the angular momenta $\bmI$ and $\bmF$ are 
quantized to the space-fixed axes;
$\bmJ$ is quantized to both the space-fixed and the body-fixed axes;
$\bmL$ and $\bmS$ are quantized to the body-fixed axes.
Without an external field, $M_F$ can be omitted:
\begin{equation}
    \ket{\state, v, \varLambda, S,\varSigma,J,\varOmega,I,F}.
    \label{eq:caseabeta}
\end{equation}
The basis functions are countable in \duo
and thus, can be simply denoted as:
    \begin{equation}
    \ket{k,J, I, F} = \ket{k,J}\ket{J, I,F},
    \label{eq:shortcaseabeta}
\end{equation}
where $k$ is a counting number for 
the basis functions associated with a given $J$.
It is an equivalent representation of Eq.\,\eqref{eq:caseabeta}
and $\ket{k,J}$ is short for Eq.\,\eqref{eq:casea}.

\begin{figure}
    \centering
    \includegraphics{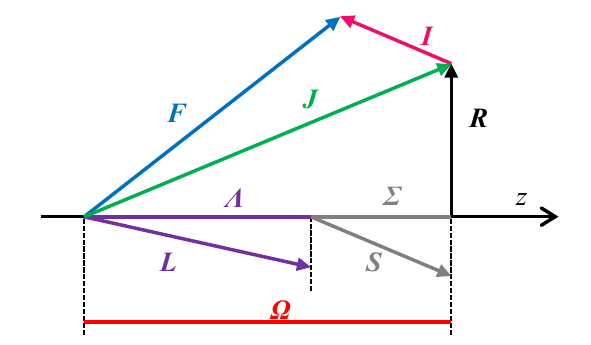}
    \caption{ Hund's case (a$_\beta$) angular momenta
    coupling scheme.
    $\bmR$ is the rotational angular momentum of
    bare nuclei.}
    \label{fig:case_a_beta}
\end{figure}

The quantum numbers,
$J$, $I$ and $F$, satisfy the triangle inequality:
\begin{equation}
    \abs*{F-I} \leq J \leq F+I.
\end{equation}
The coupling scheme used is known as Hund's
case (a$_\beta$) \cite{52FrFoxx.hyperfine}{}, and is
 illustrated in Fig.\,\ref{fig:case_a_beta}.
We emphasize that because we use complete angular basis sets, our results are independent of the coupling scheme used and
its choice largely becomes one of algorithmic convenience.
    
To obtain a parity-conserved basis set, 
we rely on the symmetrization procedure given in Eq.\,(\ref{eq:parityconserved}) by making use  
 of the eigenfunctions obtained as solutions of $\Hzero$, $\ket{\phi^{\parity,J}_m}$,
to define the  basis functions:
\begin{equation}
    \ket{\phi^{\parity,J}_m, I,F}
    =\ket{\phi^{\parity,J}_m}\ket{J,I,F}.
    \label{eq:rovibronic_basis}
\end{equation}
The parity conserved rovibronic basis
functions, Eq.\,\eqref{eq:rovibronic_basis},
can be represented by
the primitive basis functions, Eq.\,\eqref{eq:caseabeta} or Eq.\,\eqref{eq:shortcaseabeta} 
\begin{align}
    \ket{\phi^{\parity,J}_m,J,I,F}
    &=  \Identity{k,J_1}{k,J_1,I,F}\ket{\phi^{\parity,J}_m,J,I,F}\notag\\
   & = \sum_k \braket{k,J}{\phi^{\parity,J}_m}\ket{k,J, I, F},
\end{align}
where the coefficients,
$\braket{k,J}{\phi^{\parity,J}_m}$,
have been obtained when calculating 
rovibronic fine structure by solving for $\Hzero$.
The matrix elements of $\Hzero$ in this basis functions 
are straightforward
\begin{equation}
    \mel**{\phi^{\parity,J}_m, I,F}{\Hzero}{\phi^{\parity,J'}_{m'}, I,F} 
    = \kronecker{m,m'}\ \kronecker{J,J'}\ E_{m}^{\tau,J}.
    \label{eq:hzeroele}
\end{equation}
Therefore constructing the hyperfine-resolved matrix elements 
\[
    \mel**{\phi^{\parity,J}_m, I,F}{\Hzero+\Hhfs}
    {\phi^{\parity,J'}_{m'}, I,F},
\]
 just requires the matrix elements of $\Hhfs$,
\(
    \mel**{\phi^{\parity,J}_m, I,F}{\Hhfs}
    {\phi^{\parity,J'}_{m'}, I,F} .
\)

In practice, we first construct the matrix elements of $\Hhfs$ using 
 the primitive basis functions of Eq.\,(\ref{eq:caseabeta})
and then transform to the representation of $\Hzero$
of Eq.\,(\ref{eq:rovibronic_basis})
using a basis transformation.
The mathematical and physical details 
are discussed in the next two sections.
Before that,
we outline the algorithm used to calculate
hyperfine-resolved spectra.

As a first step, the hyperfine coupling curves,
such as the Fermi contact interaction  curves \cite{05FiMaWe.hyperfine}{},
are integrated over the vibrational wavefunctions.
\duo uses these vibrational matrix elements to compute
the hyperfine matrix elements
within a
Hund's case (a$_\beta$) basis set,
Eq.\,\eqref{eq:caseabeta},
and constructs a Hamiltonian matrix for each specific
total angular momentum, $F$.
Next, the matrix,
corresponding to $\Hhfs$  is constructed in the representation of Eq.~(\ref{eq:rovibronic_basis}). 
After this step,
the hyperfine matrix elements are parity conserved.
Combining the rovibronic energies and
hyperfine matrix elements,
\duo constructs the complete Hamlitonian matrix,
corresponding to $\opH$,
for each given value of $F$ and $\parity$.
Diagonalizing this matrix
gives the hyperfine-resolved energy levels and corresponding wavefunctions in the representation of Eq.~(\ref{eq:rovibronic_basis}). 
Finally, the eigenfunctions are transformed back to 
to Hund's case (a$_\beta$) representation of Eq.~\eqref{eq:caseabeta} as this representation
is more convenient to use for  hyperfine-resolved intensity calculations, for analysis of  wavefunctions
and to assign quantum number to  hyperfine states.

\section{The hyperfine structure Hamiltonian}

We investigate the field-free hyperfine structure of diatomic molecules
in which only one of the nuclei possess nuclear spin,
and consider five nuclear hyperfine terms in this work:
\begin{equation}
    \Hhfs = \HFC + \HIL + \Hdip + \HIJ + \HEQ.
\end{equation}
They are, respectively, the Hamiltonians of
the Fermi contact interaction,
the nuclear spin-orbit interaction,
the nuclear spin-electron spin dipole-dipole interaction,
the nuclear spin-rotation interaction 
and the nuclear electric quadrupole interaction.
These Hamiltonians have the following definitions
\cite{03BrCaxx.method,93Kato.methods}{}:
\begin{align}
    \HFC &= \sum_i \frac{8\uppi}{3}\, g_S\, \constantN\, 
        \DiracDelta{\bmr_{1i}}\,
          \bmI \cdot \bmS_i, \label{eq:HFCel}\\
    \HIL &= \sum_i 2\, \constantN\,
        \frac{\bmI \cdot \bmL_i}{r_{1i}^3}, \label{eq:HILel}\\
    \Hdip &= \sum_i  g_S\, \constantN\, 
      \left[\frac{\bmS_i \cdot \bmI}{r_{1i}^3}
        -\frac{3(\bmS_i\cdot\bmr_{1i})(\bmI\cdot\bmr_{1i})}{r_{1i}^5}\right], \label{eq:Hdipel}\\
    \HIJ &= c_I(R)\,\bmI \cdot \bmJ  \label{eq:HIJel},\\
    \HEQ &= \sum_{i,n}-\frac{e^2}{4\uppi\epsilon_0}\frac{r_n^2}{r_i^3}
        \sum_p (-1)^p C_p^{(2)}(\theta_i,\phi_i)
        C_{-p}^{(2)}(\theta_n,\phi_n). \label{eq:HEQel}
\end{align}
The constants,
$e$, $g_S$, $\mu_B$, $g_N$, $\mu_N$ and $\mu_0$,
are the elementary charge,
the free electron spin $g$-factor, 
the electron Bohr magneton,
the nuclear spin $g$-factor, 
the nuclear magneton
and the vacuum permeability,
respectively.
$\bmI$ is the spin of the nucleus of interest
(defined as nucleus $1$),
$\bmr_{1i}$ is the relative position between the
$i$-th electron and nucleus $1$,
$\bmS_i$ is the spin of the $i$-th electron,
$\bmL_i$ is the orbit angular momentum of the $i$-th electron,
and $\DiracDelta{\cdot}$ is the Dirac delta function.
In Eq.\,(\ref{eq:HIJel}),
we introduce the nuclear spin-rotation interaction constant, $c_I(R)$,
which is a function of internuclear distance.
Section\,8.2.2(d) of \citeauthor{03BrCaxx.method} \cite{03BrCaxx.method} 
and  \citeauthor{jt329} \cite{jt329} define
 the nuclear spin-rotation tensor
and how it can be reduced to a constant for a diatomic molecule.
In Eq.\,(\ref{eq:HEQel}),
$C^{(2)}_p$ is the modified rank-2 spherical harmonic:
\begin{equation}
    C_p^{(2)}(\theta, \phi) = \sqrt{\frac{4\uppi}{5}}\, 
    Y^{(2)}_p(\theta, \phi).
\end{equation}
where $Y^{(2)}_p(\theta, \phi)$ is the standard spherical harmonic;
$(r_i, \theta_i, \phi_i)$ and $(r_n, \theta_n, \phi_n)$
are the positions of the {$i$-th} electron
and the {$n$-th} proton, respectively.

The first four hyperfine Hamiltonians, given by Eqs.~\eqref{eq:HFCel} -- \eqref{eq:HIJel}, are nuclear magnetic dipole terms
resulting from the interactions 
between the magnetic dipole moment given by nuclear spin
and magnetic fields due to the motion of nuclei or electrons.
The nuclear electric quadrupole Hamiltonian
arises from the interaction between the nuclear electric quadrupole moment and 
the electric field inside a molecule.
The nuclear spin-rotation interaction is usually 
much weaker than the other four hyperfine terms (if non-zero).
See Table\,1 of \citeauthor{78BrViLe.hyperfine}\cite{78BrViLe.hyperfine}
for the order of magnitude of the hyperfine terms.

To aid the evaluation of matrix elements,
the hyperfine Hamiltonians can be written as
scalar products of irreducible tensor operators \cite{03BrCaxx.method}{}:
\begin{align}
    \HFC &= \sum_i \frac{8\uppi}{3}\, g_S\, \constantN\, \DiracDelta{\bmr_{1i}}\,
         \sphTensorOp{}{1}{\bmI}\cdot\sphTensorOp{}{1}{\bmS_i}, \label{eq:HFC}\\
    \HIL &= \sum_i 2\, \constantN\,\frac{1}{r_{1i}^3} \,
        \sphTensorOp{}{1}{\bmI}\cdot\sphTensorOp{}{1}{\bmL_i}, \label{eq:HIL}\\
    \Hdip &= \sum_i -\sqrt{10}\, g_S\,\constantN \,
    \sphTensorOp{}{1}{\bmI}\cdot\sphTensorOp{}{1}{\bmS_i, \bmC^{(2)}}, \label{eq:Hdip}\\
    \HIJ &= c_I(R)\, \sphTensorOp{}{1}{\bmI}\cdot\sphTensorOp{}{1}{\bmJ}, \label{eq:HIJ}\\
    \HEQ &= -e \sphTensorOp{}{2}{\grad \bmE}\cdot \sphTensorOp{}{2}{\bmQ}, \label{eq:HEQ}
\end{align}
where
$\sphTensorOp{}{k}{\cdot}$
indicates a rank-$k$ tensor.
All the tensors here
are defined in space-fixed frame.
The two tensors in Eq.\,(\ref{eq:HEQ})
defining the gradient of electric field and
the nuclear quadrupole moment are, respectively:
\begin{align}
    \sphTensorOp{}{2}{\grad \bmE}&=-\frac{1}{4\uppi\varepsilon_0}
        \sum_i \frac{e}{r_i^3}\bmC^{(2)}(\theta_i,\phi_i),\\
    e \sphTensorOp{}{2}{\bmQ} &=e
        \sum_n r_n^2 \bmC^{(2)}(\theta_n,\phi_n).
\end{align}

\section{Matrix elements of the hyperfine structure}

\subsection{Primitive matrix elements of the hyperfine structure}
In this section,
primitive matrix elements of the hyperfine structure
are initially evaluated in the representation of Eq.\,\eqref{eq:caseabeta}.
In this work,
we do not consider hyperfine couplings
between different electronic states 
when evaluating primitive matrix elements,
which are, thus, diagonal in the 
electronic state and electron spin, 
\ie, 
\[
    \state = \state',\quad  S = S',
\]
in the bra-ket notation, and immediately we have
\[
    \abs{\varLambda}= \abs{\varLambda'}. 
\]

As $\bmF = \bmJ + \bmI$,
we can initially decouple the representation of
$\ket{J, I, F, M_F}$
in Eq.\,\eqref{eq:rovibronic_basis:0} to uncoupled ones,
see \citeauthor{57Edmond.method} \cite{57Edmond.method}
for a formal definition and
irreducible spherical tensor operators.
Taking the Fermi contact term as an example,
the non-vanishing matrix element on the primitive basis functions 
for $M_F = M'_F$ is
\begin{align}
    {}&\mel**{\state, v,\varLambda,S,\varSigma,J,\varOmega,I,F}
        {\HFC}
        {\state, v',\varLambda',S,\varSigma',J',\varOmega',I,F} \notag \\
    ={}& 
        (-1)^{J'+F+I}\WignerSixj{I}{J'}{F}{J}{I}{1}
        \reducedmel{I}{\sphTensorOp{}{1}{\bmI}}{I}
        \times \frac{8\uppi}{3}\, g_S\, \constantN \notag\\
    {}&\, \times \reducedmel{\state, v,\varLambda,S,\varSigma,J,\varOmega}
            {\sum_i \DiracDelta{\bmr_{1i}}\,\sphTensorOp{}{1}{\bmS_i}}
            {\state, v',\varLambda',S,\varSigma',J',\varOmega'},
    \label{eq:decoupleHFC}
\end{align}
where $\WignerSixj{j_1}{j_2}{j_3}{j_4}{j_5}{j_6}$ is the Wigner-$6j$ symbol.
The nuclear spin is quantized to the space-fixed axes,
and thus, the reduced matrix element of 
$\sphTensorOp{}{1}{\bmI}$ is
\begin{equation}
    \reducedmel{I}{\sphTensorOp{}{1}{\bmI}}{I} 
        = \sqrt{I(I+1)(2I+1)}\,.
\end{equation}
The electron spin is quantized to the body-fixed axes.
To evaluate the second reduced matrix element in Eq.\,(\ref{eq:decoupleHFC}),
the electron spin spherical tensor is rotated 
from the space-fixed frame
to the body-fixed frame
in which the components of tensors are
denoted by $q$:
\begin{align}
    {}& \reducedmel{\state, v,\varLambda,S,\varSigma,J,\varOmega}
        {\sum_i \DiracDelta{\bmr_{1i}}\,\sphTensorOp{}{1}{\bmS_i}}
        {\state, v',\varLambda',S,\varSigma',J',\varOmega'} \notag \\
    ={}& \reducedmel{\state, v,\varLambda,S,\varSigma,J,\varOmega}
        {\sum_i \DiracDelta{\bmr_{1i}}\,\sum_q \rotateDelement{.q}{1}{\bmomega}^*\ 
            \sphTensorOp{q}{1}{\bms_i}}
        {\state, v',\varLambda',S,\varSigma',J',\varOmega'} \notag \\
    ={}& \sum_q (-1)^{J-\varOmega} 
        \WignerThreej{J}{1}{J'}{-\varOmega}{q}{\varOmega'}
        \sqrt{(2J+1)(2J'+1)}\,
        \kronecker{\varLambda,\varLambda'} \notag\\
       {} &\times \mel**{\state, v}
       {\mel**{\state,\varLambda,S,\varSigma}
        {\sum_i \DiracDelta{\bmr_{1i}}\,\sphTensorOp{q}{1}{\bms_i}}
        {\state,\varLambda',S,\varSigma'}}
       {\state, v'} , 
    \label{eq:deltaSi}
\end{align}
where $\bms_i$ is the spin of the $i$-th electron in body-fixed system,
$\rotateDelement{m',m}{k}{\bmomega}$ is a Wigner rotation matrix
and $\WignerThreej{j_1}{j_2}{j_3}{m_1}{m_2}{m_3}$ is 
a Wigner-$3j$ symbol.
The electron tensor operators,
$\sphTensorOp{q}{1}{\bms_i}$,
do not directly act on the electronic part of
Hund's case (a) basis.
We may replace the electron spin operators with 
an effective one:
\begin{align}
    &{}\frac{8\uppi}{3}\, g_S\, \constantN
    \mel**{\state, \varLambda,S,\varSigma}
        {\sum_i \DiracDelta{\bmr_{1i}}\,\sphTensorOp{q}{1}{\bms_i}}
        {\state, \varLambda,S,\varSigma'}\notag\\
    =&{}\mel**{\state,\varLambda,S,\varSigma}
        {\bF(R)\,\sphTensorOp{q}{1}{\bmS}}
        {\state,\varLambda,S,\varSigma'},
\end{align}
where $\bmS$ is the total spin.
Requiring $\varSigma = \varSigma'$, 
the Fermi contact interaction curve
can be defined as
\cite{05FiMaWe.hyperfine}{}:
\begin{equation}
    \bF(R) = \frac{8\uppi}{3}\, g_S\, \constantN
    \mel**{\state, \varLambda,S,\varSigma}
        {\sum_i \DiracDelta{\bmr_{1i}}\,
        \frac{\sphTensorOp{0}{1}{\bms_i}}{\varSigma}}
        {\state, \varLambda,S,\varSigma},
    \label{eq:bfdef}
\end{equation}
where
\(
{\sphTensorOp{0}{1}{\bms_i}}/{\varSigma} 
\) represents the projection operator
for each electron $i$
(see Eq.\,(7.152) of \citeauthor{03BrCaxx.method}\cite{03BrCaxx.method}).
Based on Eqs. \eqref{eq:decoupleHFC} to 
\eqref{eq:bfdef},
we finally get:
\begin{align}
    {}&\mel**{\state, v,\varLambda,S,\varSigma,J,\varOmega,I,F}
        {\HFC}
        {\state, v',\varLambda',S,\varSigma',J',\varOmega',I,F} \notag \\
    ={}&(-1)^{J'+F+I}\WignerSixj{I}{J'}{F}{J}{I}{1}
        \sqrt{I(I+1)(2I+1)} \notag\\
        &\times 
    \Bigg[ \sum_q (-1)^{J-\varOmega} 
        \WignerThreej{J}{1}{J'}{-\varOmega}{q}{\varOmega'}
        \sqrt{(2J+1)(2J'+1)}\,
        \notag\\
        &\qquad\times (-1)^{S-\varSigma}
        \WignerThreej{S}{1}{S}{-\varSigma}{q}{\varSigma'}
       \sqrt{S(S+1)(2S+1)}\notag\\
       &\qquad\times \kronecker{\varLambda,\varLambda'}\, 
         \mel**{\state, v}
       {\bF(R)}
       {\state, v'}
    \Bigg] \,.
\end{align}
Other hyperfine matrix elements can be evaluated analogously.

For the nuclear spin-orbit term,
we are only interested in the diagonal matrix elements
of $\varLambda$
\begin{align}
    &\mel**{\state, v,\varLambda,S,\varSigma,J,\varOmega,I,F}
        {\HIL}
        {\state, v',\varLambda,S,\varSigma',J',\varOmega',I,F}\notag\\
 ={}&  (-1)^{J'+F+I}\WignerSixj{I}{J'}{F}{J}{I}{1}
    \sqrt{I(I+1)(2I+1)} \notag\\
    &\times (-1)^{J-\varOmega} 
    \WignerThreej{J}{1}{J'}{-\varOmega}{0}{\varOmega}
    \sqrt{(2J+1)(2J'+1)}\,
    \notag\\
   {} &\times \kronecker{\varSigma,\varSigma'}\, 
    \varLambda
     \mel**{\state, v}
   {a(R)}
   {\state, v'} .
\end{align}
The non-diagonal couplings between different electronic states via  $\sphTensorOp{\pm 1}{1}{\bmL}$ are not considered here. 
The diagonal nuclear spin-orbit interaction curve
is defined as \cite{05FiMaWe.hyperfine}{}:
\begin{equation}
    a(R) = 2\, \constantN
    \mel**{\state, \varLambda, S, \varSigma}
    {\sum_i \frac{1}{r_{1i}^3}
    \frac{\sphTensorOp{0}{1}{\bml_i}}{\varLambda}}
    {\state, \varLambda, S, \varSigma},
\end{equation}
where $\bml_i$ is the orbital angular momentum of 
the $i$-th electron defined in the body-fixed frame.

The nuclear spin-electron spin dipole-dipole
interaction is somewhat complicated.
With the definition
(see Appendix 8.2 of \citeauthor{03BrCaxx.method}
\cite{03BrCaxx.method}):
\begin{equation}
    \sphTensorOp{q}{1}{\bms_i, \bmC^{(2)}} 
    = 
     \sum_{q_1,q_2} -\sqrt{3}(-1)^q\, \sphTensorOp{q_1}{1}{\bms_i}\,
     \frac{C^{(2)}_{q_2}(\theta_{1i},\phi_{1i})}{r_{1i}^3}\,     
     \WignerThreej{1}{2}{1}{q_1}{q_2}{-q},
\end{equation}
where $(r_{1i}, \theta_{1,i}, \phi_{1,i})$
are the spherical polar coordinates of electron $i$ relative to
nucleus 1,
we shall give two kinds of matrix elements.
For the term diagonal in $\varLambda$,
\ie $q_2 = 0$ and $q = q_1$:

\begin{align}
    &\mel**{\state, v,\varLambda,S,\varSigma,J,\varOmega,I,F}
    {\Hdip}
    {\state, v',\varLambda,S,\varSigma',J',\varOmega',I,F}\notag\\
   ={}& (-1)^{J'+F+I}\WignerSixj{I}{J'}{F}{J}{I}{1}
    \sqrt{I(I+1)(2I+1)}\notag\\
     & \times  \Bigg[\sum_{q}\,
     (-1)^{J-\varOmega} 
    \WignerThreej{J}{1}{J'}{-\varOmega}{q}{\varOmega'}
    \sqrt{(2J+1)(2J'+1)}\notag \\
    &\,\qquad\times
    (-1)^{q} \sqrt{30} \WignerThreej{1}{2}{1}{q}{0}{-q}
    (-1)^{S-\varSigma}
    \WignerThreej{S}{1}{S}{-\varSigma}{q}{\varSigma'}
    \sqrt{S(S+1)(2S+1)}
    \notag \\
    &\qquad\times\frac{1}{3}
    \mel**{\state, v}{ \,c(R)}{\state, v'}
    \Bigg], 
\end{align}
The diagonal 
nuclear spin-electron spin dipole dipole interaction
constant curve is defined as
\cite{05FiMaWe.hyperfine}{},
\begin{equation}
        c(R) = 3\, g_S\,\constantN
        \mel**{\state, \varLambda, S, \varSigma}
        {\sum_i 
        \frac{C_{0}^{(2)}(\theta_{1i},\phi_{1i})}{r_{1i}^3}
        \frac{\sphTensorOp{0}{1}{\bms_i}}{\varSigma}}
        {\state, \varLambda, S, \varSigma}.
        \label{eq:dipolarc} \\
\end{equation}
For the off-diagonal terms of $\Hdip $ in $\varLambda$ and $\varLambda'$
which satisfy $q_2=\mp 2$,
\ie $q_1 = \pm 1$ and $q=\mp 1$,  we have
\begin{align}
    &\mel**{\state, v,\varLambda,S,\varSigma,J,\varOmega,I,F}
        {\Hdip}
        {\state, v',\varLambda',S,\varSigma',J',\varOmega',I,F}\notag\\
   ={}& (-1)^{J'+F+I}\WignerSixj{I}{J'}{F}{J}{I}{1}
    \sqrt{I(I+1)(2I+1)}\notag\\
     & \times (-1)^{J-\varOmega} 
    \WignerThreej{J}{1}{J'}{-\varOmega}{\mp 1}{\varOmega'}
    \sqrt{(2J+1)(2J'+1)}\notag \\
    &\, \times 
    \frac{\sqrt{S(S+1)-\varSigma(\varSigma\pm 1)}}{\mp \sqrt{2}}
    \mel**{\state, v}{d(R)}{\state, v'},
\end{align}
The off-diagonal 
nuclear spin-electron spin dipole dipole interaction
constant curve is defined as
\cite{05FiMaWe.hyperfine}{},
\begin{align}
    d(R) &= -\sqrt{6}\, g_S\,\constantN \notag \\
    &\times
    \mel**{\state, \varLambda, S, \varSigma}
    {\sum_i 
    \frac{C_{\mp 2}^{(2)}(\theta_{1i},\phi_{1i})}{r_{1i}^3}
    \frac{\mp\sqrt{2}\,\sphTensorOp{\pm 1}{1}{\bms_i}}{ \sqrt{S(S+1)-\varSigma(\varSigma\pm 1)}}
    }
    {\state, \varLambda', S, \varSigma'}.
    \label{eq:dipolard}
\end{align}

The case of the nuclear spin-rotation 
interaction is much simpler,
as it is not necessary to rotate $\sphTensorOp{}{1}{\bmJ}$
to the body-fixed axis system:
\begin{align}
    &\mel**{\state, v,\varLambda,S,\varSigma,J,\varOmega,I,F}
        {\HIJ}
        {\state, v',\varLambda',S,\varSigma',J',\varOmega',I,F}\notag\\
 ={}&   (-1)^{J'+F+I}\WignerSixj{I}{J'}{F}{J}{I}{1}
        \sqrt{I(I+1)(2I+1)} \notag\\
        &\times (-1)^{J-\varOmega} 
        \WignerThreej{J}{1}{J'}{-\varOmega}{0}{\varOmega}
        \kronecker{\varLambda,\varLambda'}\,
     \kronecker{\varSigma,\varSigma'}\, 
     \kronecker{J,J'}\sqrt{J(J+1)(2J+1)}\,\notag\\
    &\times\mel**{\state, v}{c_I(R)}{\state, v'}.
\end{align}

To evaluate the matrix elements for the electric quadrupole interaction,
we decouple the inner product
of second rank irreducible tensors:
\begin{align}
    {}&\mel**{\state, v,\varLambda,S,\varSigma,J,\varOmega,I,F}
        {\HEQ}
        {\state, v',\varLambda',S,\varSigma',J',\varOmega',I,F} \notag \\
    ={}& (-1)^{J'+I+F}\WignerSixj{I}{J}{F}{J'}{I}{2} 
        \reducedmel{I}{-e\sphTensorOp{}{2}{\bmQ}}{I}\notag\\
     {}&\times   \reducedmel{\state, v,\varLambda,S,\varSigma,J,\varOmega}
            {\sphTensorOp{}{2}{\grad \bmE}}
            {\state, v',\varLambda',S,\varSigma',J',\varOmega'}.
\end{align}
The electric quadrupole reduced matrix element
is non-zero only if $I\ge 1$; it
can be evaluated as
\begin{align}
    {}&\reducedmel{I}{-e\sphTensorOp{}{2}{\bmQ}}{I}
        = \frac{-eQ}{2}\WignerThreej{I}{2}{I}{-I}{0}{I}^{-1}.
\end{align}
where $eQ$ is
the nuclear electric quadrupole moment, 
see \citeauthor{71CoLuxx.hyperfine}\cite{71CoLuxx.hyperfine} or
Appendix 8.4 of \citeauthor{03BrCaxx.method}
\cite{03BrCaxx.method}{}.
The values of $Q$ for various atoms
were collected by \citeauthor{18Pyxxxx.hyperfine}\cite{18Pyxxxx.hyperfine}{}.
The reduced matrix element of the gradient of  electric field is
\begin{align}
    {}&\reducedmel{\state, v,\varLambda,S,\varSigma,J,\varOmega}
        { \sphTensorOp{}{2}{\grad \bmE}}
        {\state, v',\varLambda',S,\varSigma',J',\varOmega'} \notag \\
    ={}& \sum_q (-1)^{J-\varOmega}
        \WignerThreej{J}{2}{J'}{-\varOmega}{q}{\varOmega'}
        \sqrt{(2J+1)(2J'+1)}\,
        \kronecker{\varSigma,\varSigma'}\notag\\
    {}&\times
        \mel**{\state, v}{
        \mel**{\state, \varLambda, S,\varSigma}{\sphTensorOp{q}{2}{\grad \bmE}}{\state, \varLambda',S,\varSigma'}
        }{\state, v'}.
\end{align}
The diagonal and off-diagonal $R$-dependent constants of the gradient of 
electric field are
respectively defined as (see 
Eqs. (7.159) and (7.163) of
\citeauthor{03BrCaxx.method}\cite{03BrCaxx.method}):
\begin{align}
    q_0(R)&=
    -2\mel**{\state, \varLambda, S, \Sigma}
        {\sphTensorOp{0}{2}{\grad \bmE}}
        {\state, \varLambda, S, \Sigma},
        \label{eq:eqq0}\\
    q_2(R)&=
        -2\sqrt{6}\mel**{\state, \varLambda, S, \Sigma}
        {\sphTensorOp{\pm 2}{2}{\grad \bmE}}
        {\state, \varLambda', S, \Sigma}.
    \label{eq:eqq2}
\end{align}
Note that sometimes 
$q_0$ is denoted as $q_1$, see \eg Eq.\,(2.3.76 a) of \citeauthor{85Hixxxx.method}\cite{85Hixxxx.method}{}.
We follows the convention of \citeauthor{03BrCaxx.method}\cite{03BrCaxx.method}
and preserve the variable $q_1$ for 
the nuclear electric quadrupole coupling constant between different electronic states arising from
${\sphTensorOp{\pm 1}{2}{\grad \bmE}}$
which will be the subject of  future work.
Finally, the diagonal matrix elements of
nuclear electric quadrupole coupling are
\begin{align}
    {}&\mel**{\state, v,\varLambda,S,\varSigma,J,\varOmega,I,F}
        {\HEQ}
        {\state, v',\varLambda',S,\varSigma',J',\varOmega',I,F} \notag \\
    ={}& (-1)^{J'+I+F}\WignerSixj{I}{J}{F}{J'}{I}{2} 
        \WignerThreej{I}{2}{I}{-I}{0}{I}^{-1}\notag\\
     {}&\times   (-1)^{J-\varOmega}
        \WignerThreej{J}{2}{J'}{-\varOmega}{0}{\varOmega}
        \sqrt{(2J+1)(2J'+1)}\,
        \kronecker{\varSigma,\varSigma'}\,
    \kronecker{\varLambda, \varLambda'}\,
       \notag\\
    {}&\times 
        \frac{1}{4}\mel**{\state, v}{eQq_0(R)}{\state, v'},
\end{align}
while the off-diagonal ones are:
\begin{align}
    {}&\mel**{\state, v,\varLambda,S,\varSigma,J,\varOmega,I,F}
        {\HEQ}
        {\state, v',\varLambda',S,\varSigma',J',\varOmega',I,F} \notag \\
    ={}& (-1)^{J'+I+F}\WignerSixj{I}{J}{F}{J'}{I}{2} 
        \WignerThreej{I}{2}{I}{-I}{0}{I}^{-1}\notag\\
     {}&\times   (-1)^{J-\varOmega}
        \WignerThreej{J}{2}{J'}{-\varOmega}{\pm 2}{\varOmega'}
        \sqrt{(2J+1)(2J'+1)}\,
        \kronecker{\varSigma,\varSigma'}\notag\\
    {}&\times\frac{1}{4\sqrt{6}}
        \mel**{\state, v}{eQq_2(R)}{\state, v'}.
\end{align}

As we only consider the hyperfine interactions 
within a particular electronic state in this paper,
the off-diagonal matrix elements arising from $d(R)$ in Eq.\,\eqref{eq:dipolard}
and $q_2(R)$ in Eq.\,\eqref{eq:eqq2}
only contribute to the \Ldoubling terms of $\Pi$ states.
In the electron spin resonance spectroscopy literature,
the Fermi-contact and nuclear spin-electron spin dipole-dipole terms
are, respectively, the first-order isotropic 
and dipolar contributions to 
the hyperfine coupling $\bm{A}$-tensor
\cite{21FeDuAu.hyperfine}.
When the second-order contributions of
paramagnetic spin orbit (PSO) interaction
are considered,
the hyperfine coupling constants defined
in this paper
can be further revised 
by the PSO terms
and determined by
the matrix elements of total hyperfine $\bm{A}$-tensor
\cite{04ArVaKa.hyperfine}{}.

\subsection{Parity conserved matrix elements
under the rovibronic wavefunctions}

Recall the short notation of  
Hund's case (a$_\beta$) basis in 
Eq.\,(\ref{eq:shortcaseabeta}),
$\ket{k,J, I, F}$ 
and the 
basis functions we defined in Eq.\,(\ref{eq:rovibronic_basis}),
\(
    \ket{\phi^{\parity,J}_m, I,F},
\)
the hyperfine matrix elements under the basis set can be expanded as
\begin{align}
    &{} \mel**{\phi^{\parity,J}_m,I,F}{\Hhfs}{\phi^{\parity,J'}_{m'},I,F} \notag\\
    = &{}
         \mel**{\phi^{\parity,J}_m,I,F}{\Identity{k,J_1}{k,J_1,I,F} \Hhfs 
        \Identity{k',J_2}{k',J_2,I,F}}{\phi^{\parity,J'}_{m'},I,F} \notag\\
    = &{}\sum_{k,J_1} \sum_{k',J_2}
        \braket{\phi^{\parity,J}_m,I,F}{k,J_1,I,F} 
        \mel**{k,J_1,I,F}{\Hhfs}{k',J_2,I,F} 
        \braket{k',J_2,I,F}{\phi^{\parity,J'}_{m'},I,F} \notag \\
    = &{}\sum_{k} \sum_{k'}
        \braket{\phi^{\parity,J}_m,I,F}{k,J,I,F} 
        \mel**{k,J,I,F}{\Hhfs}{k',J',I,F} 
        \braket{k',J',I,F}{\phi^{\parity,J'}_{m'},I,F}.
    \label{eq:transformstart}
\end{align}
We can rewrite the basis transformation
into the matrix format:
\begin{equation}
    \mat{H}^{\parity, F}_\mathrm{hfs} =  
        \left(\mat{\varPhi}^{\parity, F}\right)^\dagger\  
        \mat{H}^{F}_\mathrm{hfs}\ \mat{\varPhi}^{\parity, F}.
\end{equation}
\(\mel**{\phi^{\parity,J}_m,I,F}{\Hhfs}{\phi^{\parity,J'}_{m'},I,F}\),
\(\mel**{k,J,I,F}{\Hhfs}{k',J',I,F}\) and 
\(\braket{k',J',I,F}{\phi^{\parity,J'}_{m'},I,F}\) 
are the matrix elements of \( \mat{H}^{\parity, F}_\mathrm{hfs}\),
\(\mat{H}^{F}_\mathrm{hfs}\) and 
\(\mat{\varPhi}^{\parity, F}\), respectively,
and,
\[
    \braket{k',J',I,F}{\phi^{\parity,J'}_{m'},I,F} =
    \braket{k',J'}{\phi^{\parity,J'}_{m'}}\,.
\]

\subsection{Solution for the hyperfine structure}

The final Hamiltonian which is constructed 
from  summation of the rovibronic and 
hyperfine matrices
\begin{equation}
    \mat{H}^{\parity, F} =
        \mat{H}^{(0),\parity, F} +
        \mat{H}^{\parity, F}_\mathrm{hfs},
\end{equation}
where $\mat{H}^{(0),\parity, F}$ is the matrix of $\Hzero$
(see Eq.\,\eqref{eq:hzeroele}
for the matrix elements).
Diagonalizing the parity-conserved matrix of each $F$
results in the energies and wavefunctions of 
hyperfine structure:
\begin{equation}
    \mat{E^}{\parity, F} =
    \left(\mat{U}^{\parity, F}\right)^\dagger\,
    \mat{H}^{\parity, F} \,
    \mat{U}^{\parity, F} \,.
\end{equation}
The eigenfunction matrix $\mat{U}^{\parity, F}$
is represented in the parity-conserved rovibronic basis set defined in 
Eq.\,(\ref{eq:rovibronic_basis}),
which is, however, not very useful for
quantum number assignments and 
wavefunction analysis.
For these purposes,
the wavefunctions can be transformed back 
in the representation of Hund's case (a) basis set
and the final wavefunction matrix is 
\begin{equation}
    \mat{\varPsi}^{\parity, F} =
    \mat{\varPhi}^{\parity, F}\,
    \mat{U}^{\parity, F}\, .
    \label{eq:transformend}
\end{equation}
Here, 
we denote the countable
rovibronic wavefunctions considering nuclear hyperfine interaction as 
\begin{equation}
    \ket{\psi_m^{\parity, F}},
    \label{eq:hyperwave}
\end{equation}
such that
\begin{equation}
    \mel**{\psi_m^{\parity, F}}{\opH}{\psi_{m'}^{\parity, F}}
    = \kronecker{m,m'}\,E_m^{\parity, F},
\end{equation}
where $E_m^{\parity, F}$ is the corresponding eigenvalue of $\ket{\psi_m^{\parity, F}}$.

The basis transformation procedures from Eq.\,(\ref{eq:transformstart})
to Eq.\,(\ref{eq:transformend})
reveal the key feature of our variational method which involves accounting for
the contribution of every  basis function
to the final eigenstates. 
Finally,
only $F$, $\parity$, and counting number $m$
are good quantum numbers.

\section{Line strength of the hyperfine transitions}
In the absence of an external field,
the line strength of a nuclear spin resolved rovibronic transition is defined by\cite{71CoLuxx.hyperfine}
\begin{align}
    &S(m, \parity, F \leftrightarrow m', \parity', F')\notag\\
    =& \sum_{p,M_F, M_{F'}}
        \left|\mel**{\psi_m^{\parity, F},M_F}
        {\sphTensorOp{p}{1}{\bmmutrans}}
        {\psi_{m'}^{\parity',F'},M_{F'}}\right|^2 \notag\\
    = {}&\left|\reducedmel{\psi_m^{\parity, F}}
    {\sphTensorOp{\,}{1}{\bmmutrans}}
    {\psi_{m'}^{\parity',F'}}\right|^2
    \left[
        \sum_{p,M_F, M_{F'}}
        \left|\WignerThreej{F}{1}{F'}{-M_F}{p}{M_{F'}}
        \right|^2  
    \right] \notag\\
    ={}&\left|\reducedmel{\psi_m^{\parity, F}}
    {\sphTensorOp{}{1}{\bmmutrans}}
    {\psi_{m'}^{\parity',F'}}\right|^2
    \label{eq:linestrength}
\end{align}
We initially evaluate the reduced matrix elements of the
electric dipole moment in the representation 
of Eq.\,(\ref{eq:caseabeta})  and 
then calculate the reduced line strength matrix elements by 
matrix multiplication:
\begin{equation}
    ^{\parity, F}\mat{D}^{\parity', F'} =  
    (\mat{\varPsi}^{\parity, F})^\dagger\,  
    ^{F}\mat{D}^{F'}\, 
    \mat{\varPsi}^{\parity', F'},
\end{equation}
where $^{F}\mat{D}^{F'}$ 
and $^{\parity, F}\mat{D}^{\parity', F'}$ are the 
reduced transition dipole moment matrices in the representation 
of Eq.\,(\ref{eq:caseabeta}) and Eq.\,(\ref{eq:hyperwave}) 
respectively.
The following equations give 
the elements of $^{F}\mat{D}^{F'}$,
\ie 
\(
\reducedmel{k,J,I,F}{\sphTensorOp{}{1}{\bmmu}}{k',J',I,F}
\).

As $\bmF = \bmJ + \bmI$
and $\sphTensorOp{}{1}{\bmmutrans}$ commutes with $\bmI$,
\begin{align}
    {}&\reducedmel{\state, v,\varLambda,S,\varSigma,J,\varOmega, M_J,I,F}
        {\sphTensorOp{}{1}{\bmmutrans}}
        {\state', v',\varLambda',S',\varSigma',J',\varOmega', M_{J'},I,F'} \notag\\
    ={}& (-1)^{J+I+F'+1}
    \sqrt{(2F+1)(2F'+1)}
    \WignerSixj{J'}{F'}{I}{F}{J}{1} \notag\\
    {}&\quad\times
    \reducedmel{\state, v,\varLambda,S,\varSigma,J,\varOmega}
    {\sphTensorOp{}{k}{\bmmutrans}}
    {\state', v',\varLambda',S',\varSigma',J',\varOmega'}.
    \label{eq:dipolematele}
\end{align}
Rotating the spherical tensor to the body-fixed frame gives:
\begin{align}
    &{}\reducedmel{\state, v,\varLambda,S,\varSigma,J,\varOmega}
    {\sphTensorOp{}{k}{\bmmutrans}}
    {\state', v',\varLambda',S',\varSigma',J',\varOmega'} \notag \\
    = &{} \reducedmel{\state, v,\varLambda,S,\varSigma,J,\varOmega}
    {\sum_{q=-1}^{1}\rotateDelement{.q}{1}{\bmomega}^*\ \sphTensorOp{q}{1}{\bmmutrans}}
    {\state', v',\varLambda',S',\varSigma',J',\varOmega'}\notag\\
    =&{} \sum_{q=-1}^{1} (-1)^{J-\varOmega} \sqrt{(2J+1)(2J'+1)}
    \WignerThreej{J}{1}{J'}{-\varOmega}{q}{\varOmega'} \notag\\
    &\qquad \times \mel**{\state, v,\varLambda,S,\varSigma}
    {\sphTensorOp{q}{1}{\bmmutrans}}
    {\state', v',\varLambda',S',\varSigma'}.
\end{align}
The matrix element 
\(\mel**{\state, v,\varLambda,S,\varSigma}
{\sphTensorOp{q}{1}{\bmmutrans}}
{\state', v',\varLambda',S',\varSigma'}\)
is the same as the one used for 
the calculation of rovibronic transition intensities
excluding nuclear spin
in \duo \cite{jt609}{},
\begin{equation}
    \mel**{\state, v,\varLambda,S,\varSigma}
        {\sphTensorOp{q}{1}{\bmmutrans}}
        {\state', v',\varLambda',S',\varSigma'}
    =  \mel**{\state, v}
    {\mu_q(R)}
    {\state', v'},
\end{equation}
where $\mu_q(R)$ is
the electric dipole moment curve
represented in the body-fixed frame which
can be obtained from \abinitio calculation.

For dipole moment transitions
parity has to be changed
and thus follows the selection rule:
\begin{equation}
   \parity: \quad + \Leftrightarrow - 
\end{equation}
The selection rules on $F$ comes from the Wigner-$6j$
symbol of Eq.\,(\ref{eq:dipolematele}):
\begin{equation}
    \Delta F = -1, 0, 1;\ \text{and}\  F \neq 0\ \text{if}\ \Delta F = 0.
\end{equation}

The hyperfine Hamiltonian mixes wavefunctions
with different $J$; as a result,
electric dipole transition
`forbidden' lines with $|\Delta J| > 1$ are observable.
For example, when $I=1/2$,
we can observe electric dipole transitions of $O$ and $S$ branches
($\Delta J = \pm 2$), 
even if they might be much weaker
than the transitions of $P$, $Q$ and $R$ branches.

\section{Numerical verification}
To illustrate and validate our new hyperfine  modules,
we calculate hyperfine-resolved rotational spectra for electronic and vibrational ground state of \ce{^{14}N^{16}O} and \ce{^{24}Mg^{1}H}.
While both \ce{^{16}O} and \ce{^{24}Mg} have nuclear spin zero; \ce{^{14}N}  has $I=1$ and \ce{^{1}H} has $I=1/2$ which allows
us to test different coupling mechanisms. For this purpose we compare the results of our \duo\ calculations with 
 of PGOPHER \cite{17Wester} using the same model for each calculation.
PGOPHER obtains the energy levels and spectra from effective Hamiltonians given appropriate 
 spectral constants. In contrast,
\duo takes in coupling curves and performs variational calculations.
To get consistent inputs between the two codes it was necessary to simplify the treatment used by \duo.

For \ce{^{14}N^{16}O} we approximate the \duo\ solution by using  only one contracted vibrational basis function, \ie $\ket{\XPi, v=0}$ which ensures that we avoid any hyperfine-induced interaction between different vibrational states. 
In PGOPHER, we used values for the 
rotational constant, $B_0$,
and spin-orbit coupling constant matrix, $A_0$,  computed using \duo:
\begin{align}
    B_0 &= \mel**{\XPi, v=0}{\frac{\hbar^2}{2\mu R^2}}{\XPi, v=0},\\
    A_0 &= 2\mel**{\XPi, v=0}{C_{\mathrm{SO}}(R)}{\XPi, v=0},
\end{align}
where $\mu$ is the reduced mass of \ce{^{14}N^{16}O}
and $C_{\mathrm{SO}}(R)$ is the spin-orbit coupling curve.
Note that,
for spin-orbit interaction,
the coupling curve, $C_\mathrm{SO}(R)$, describes the coupling energies,
while the constant, $A$,
is defined by the splitting energies.
Thus, $A$ is defined by twice the matrix element.
The NO X $^2\Pi$ potential energy curve used by \duo\ was taken from  
Wong \etal \cite{17WoYuBe.NO}{}. 
$C_{\mathrm{SO}}(R)$ was assigned an artificial constant
$C_{\mathrm{SO}}(R) = \SI{60}{\per\cm}$ 
and the transition dipole moment curve was set to \SI{1}{Debye}.
Our adopted values for $B_0$ and $A_0$ are given in Table\,\ref{tab:NOconstant}.

For this analysis, the hyperfine coupling  was chosen using an artificial curves  much greater than experimental values.
By including only one hyperfine constant at a time,
we test the affects of a particular hyperfine interaction.
The results are compared in Table\,\ref{tab:NOcompare}.
Note that, PGOPHER uses nuclear spin-electron spin constants, $b$,
defined by Frosch and Foley \cite{52FrFoxx.hyperfine}{},
rather than $\bF$. They are related by the dipole-dipole constant, $c$,
\begin{equation}
    \bF = b + \frac{c}{3}.
\end{equation}
\duo achieves excellent agreement with PGOPHER
for the calculation of both the line positions $\nu$ and line
strengths $S$.
The slight differences are due to rounding error.
As we did not include \Ldoubling terms in our calculation
the wavenumbers corresponding to $\bF$, $a$, $eQq_0$ and $c_I$
in the first and second columns of the same $F=0.5$ 
(or in the third and fourth columns, $F=1.5$) of Table\,\ref{tab:NOcompare}
are the same.
Hyperfine interactions only splits the transitions of different $F$
in the first and third columns (or in the second and fourth columns).
In contrast,
the wavenumbers obtained with $eQq_2$ or $d$ included  are different from each other even for the same values of $F$
due to the hyperfine contribution to both \Ldoubling and hyperfine splitting.

\begin{table}
\centering
    \caption{Spectroscopic constants for \ce{^{14}N^{16}O} used in this paper.}
    \label{tab:NOconstant}
    \begin{tabular}{ll}
        \hline
        Constants & Values [\si{\per\cm}]\\
        \hline
        $B_0$ & \num{1.69608401191395}\\
        $A_0$ & \num{120}\\
        \hline
    \end{tabular}
\end{table}

\begin{table}
    \centering
    \caption{Comparison of \ce{^{14}N^{16}O} line positions and line strengths
    for
    calculated results from \duo and PGOPHER.
    Hyperfine constants are in \si{\per\cm} and 
    line positions are given in \si{\MHz}.
    The line strength, $S$
    [\si{Debye^2}],
    has the same definition as that in PGOPHER
    when the intensity unit option of PGOPHER,
    \texttt{IntensityUnit}, is chosen 
    as \texttt{HonlLondon}
    and the transition dipole moment
    is set to \SI{1}{Debye}.}
    \label{tab:NOcompare}
      \begin{tabular}{cr|rrrr}
      \hline
      \multicolumn{2}{c|}{Number} & 1     & 2     & 3     & 4     \\
      \hline
      \multirow{3}[2]{*}{Upper} & $F'$   & 0.5   & 0.5   & 1.5   & 1.5    \\
            & $\parity''$  & $-$     & $+$     & $-$     & $+$      \\
            & $J'' $    & 1.5   & 1.5   & 1.5   & 1.5    \\
      \hline
      \multirow{3}[2]{*}{Lower} & $F''$  & 0.5   & 0.5   & 0.5   & 0.5   \\
            & $\parity''$   & $+$     & $-$     & $+$     & $-$     \\
            & $J''$    & 0.5   & 0.5   & 0.5   & 0.5    \\
      \hline
      \multirow{4}[2]{*}{$b=0.1$ $c=0.3$} & $\nu_\duo$ & 148343.21846  & 148343.21846  & 147225.55589  & 147225.55589  \\
            & $\nu_\text{PG}$ & 148343.21850  & 148343.21850  & 147225.55590  & 147225.55590    \\
            & $S_\duo$ & 0.60757296  & 0.60757296  & 0.77125182  & 0.77125182    \\
            & $S_\text{PG}$    & 0.60757300  & 0.60757300  & 0.77125180  & 0.77125180   \\
      \hline
      \multirow{4}[2]{*}{$a=0.1$} & $\nu_\duo$ & 151349.03162  & 151349.03162  & 151956.77196  & 151956.77196  \\
            & $\nu_\text{PG}$  & 151349.03160  & 151349.03160  & 151956.77200  & 151956.77200   \\
            & $S_\duo$ & 0.58421238  & 0.58421238  & 0.72433238  & 0.72433238   \\
            & $S_\text{PG}$  & 0.58421240  & 0.58421240  & 0.72433240  & 0.72433240   \\
      \hline
      \multirow{4}[2]{*}{$eQq_0=0.1$} & $\nu_\duo$ & 149591.09156  & 149591.09156  & 150930.88155  & 150930.88155  \\
            & $\nu_\text{PG}$  & 149591.09160  & 149591.09160  & 150930.88160  & 150930.88160  \\
            & $S_\duo$ & 0.59805081  & 0.59805081  & 0.73432902  & 0.73432902   \\
            & $S_\text{PG}$  & 0.59805080  & 0.59805080  & 0.73432900  & 0.73432900   \\
      \hline
      \multirow{4}[2]{*}{$c_I=0.1$} & $\nu_\duo$ & 145827.72503  & 145827.72503  & 150324.61190  & 150324.61190  \\
            & $\nu_\text{PG}$  & 145827.72500  & 145827.72500  & 150324.61190  & 150324.61190   \\
            & $S_\duo$ & 0.59221720  & 0.59221720  & 0.74027149  & 0.74027149  \\
            & $S_\text{PG}$  & 0.59221720  & 0.59221720  & 0.74027150  & 0.74027150   \\
      \hline
      \multirow{4}[2]{*}{$eQq_2=0.1$} & $\nu_\duo$ & 150346.43930  & 150302.88914  & 150307.21201  & 150342.05212    \\
            & $\nu_\text{PG}$  & 150346.43930  & 150302.88910  & 150307.21200  & 150342.05210    \\
            & $S_\duo$ & 0.59221687  & 0.59221668  & 0.74027121  & 0.74027140    \\
            & $S_\text{PG}$  & 0.59221690  & 0.59221670  & 0.74027120  & 0.74027140   \\
      \hline
      \multirow{4}[2]{*}{$d=0.1$} & $\nu_\duo$ & 150329.98859  & 150332.52077  & 149133.39987  & 151532.62042    \\
            & $\nu_\text{PG}$  & 150329.98860  & 150332.52080  & 149133.39990  & 151532.62040   \\
            & $S_\duo$ & 0.59210956  & 0.59211520  & 0.75214574  & 0.72851989   \\
            & $S_\text{PG}$  & 0.59210960  & 0.59211520  & 0.75214570  & 0.72851990    \\
      \hline
      \end{tabular}
  \end{table}

We also tested the code for an $I=1/2$ case by
calculating  pure rotational transitions within the 
$v=0$, $\mathrm{X}\,^2\Sigma^+$ state of \ce{^{24}MgH}, again 
using a unit electric dipole moment curve.
This is a rather realistic case,
as the input spectral constants to PGOPHER
listed in Table\,\ref{tab:MgHconstants}
were determined by the observed transitions\cite{93ZiBaAn.MgH}{}.
As for the input to \duo,
the potential energy curve was shifted from an empirically-determined one \cite{jt529,jt857}
to reproduce the $B_0$ constant given in Table\,\ref{tab:MgHconstants}, \ie
\begin{equation}
    B_0 = \mel**{\XSigma, v=0}{\frac{\hbar^2}{2\mu R^2}}{\XSigma, v=0}
\end{equation}
The curves of spin-rotation and hyperfine couplings were defined as:
\begin{align}
    \gamma(R) &= \gamma_0,\\
    \bF(R) &= b_0 + \frac{c_0}{3},\\
    c(R) & = c_0.
\end{align}
Note that the contribution of $D_0$ is not allowed for
when only one contracted basis function is used in \duo.
Just like the $B_v$ constant,
\duo does not use rotational constants, 
$D_v$, $H_v$, \etc, either and introduction of these
centrifugal distortion would require 
manipulation of the potential energy curves 
which are beyond the scope of this work.
Nevertheless,
\duo still gives hyperfine splittings which are consistent  with PGOPHER,
see the  comparison in Table\,\ref{tab:mghsplitsv1},
because $D_0$ uniformly shifts the hyperfine energy levels
within the same $N$ rotational levels, where
$N$ is the quantum number corresponding to $\bm N$ which is defined as:
\begin{equation}
    \bm{N} = \bmJ -\bmS.
\end{equation}

\begin{table}[htb]
    \caption{
    $\XSigma$, $v=0$ spectral constants of \ce{^{24}Mg^{1}H} 
    determined by \citeauthor{93ZiBaAn.MgH}\cite{93ZiBaAn.MgH}
    These values were used as the input to PGOPHER.
    }
    \label{tab:MgHconstants}
    \centering
    \begin{tabular}{r|r}
        \hline
        Constants &  Values [\si{MHz}]\\
        \hline
         $B_0$ & \num{171976.1782}\\
         $D_0$ & \num{10.6212}\\
         $\gamma_0$ & \num{790.809}\\
         $b_0$ & \num{306.277}\\
         $c_0$ & \num{4.792}\\
         \hline
    \end{tabular}
\end{table}

\begin{table}
\caption{Comparison of \ce{^{24}Mg^1H} $\XSigma, v=0$
hyperfine energies calculated by \duo and PGOPHER.
Only one vibrational contracted basis function $\ket{\XSigma,v=0}$ was
used in this case.
All energies are given in MHz.}
\label{tab:mghsplitsv1}
\begin{tabular}{r|rrr|r|rrr}
\hline
No. & $F$     & $\parity$ & $J$     & $N$     & $E_\duo$ & $E_\mathrm{PG}$ & Difference \\
\hline
1     & 0     & $+$    & 0.5   & 0 & -230.9057 & -230.9057 & 0.0000 \\
2     & 1     & $+$     & 0.5   &0       & 76.9686 & 76.9686 & 0.0000 \\
\hline
3     & 1     & $-$     & 0.5   & 1 & 343117.2196 & 343074.7347 & 42.4849 \\
4     & 0     & $-$     & 0.5   &1       & 343236.9188 & 343194.4339 & 42.4849 \\
5     & 1     & $-$     & 1.5   & 1      & 344238.9505 & 344196.4655 & 42.4850 \\
6     & 2     &$-$     & 1.5   & 1      & 344424.5699 & 344382.0849 & 42.4850 \\
\hline
\end{tabular}
\end{table}

We then allowed for the effect of vibrational coupling in \duo by
increasing  contracted vibration bases was set to five functions,
\ie, $\ket{\XSigma, v=0, 1, 2, 3, 4}$. 
As shown in Table\,\ref{tab:mghsplitsv5},
vibrational coupling from higher vibrational states automatically
introduces centrifugal distortion to the $v=0$ state 
and improves the accuracy of the calculation,
compared with the lower rotational levels in Table\,\ref{tab:mghsplitsv1}.
We did not use a very accurate model here,
and thus for higher rotational levels,
we still got
obvious energy differences
in Table\,\ref{tab:mghsplitsv5},
and frequency differences in Table\,\ref{tab:mghtransitionsv5}.
The best way to achieve experimental accuracy 
is to refine the curves by fitting calculated energies or frequencies
to measured ones.

\begin{table}
    \centering
    \caption{Comparison of \ce{^{24}Mg^1H} $\XSigma, v=0$
hyperfine energies calculated by \duo and PGOPHER.
Five vibrational contracted basis functions $\ket{\XSigma,v=0, 1, 2, 3, 4}$ were
        used in this case.
All energies are given in MHz.}
    \label{tab:mghsplitsv5}
\begin{tabular}{r|rrr|r|rrr}
\hline
No.   & $F$     & $\parity$ & $J$     & $N$     & $E_\duo$  & $E_\mathrm{PG}$   &  Difference  \\
\hline
1     & 0     & $+$     & 0.5   & 0     & -230.9058 & -230.9057 & -0.0001 \\
2     & 1     & $+$    & 0.5   & 0     & 76.9686 & 76.9686 & 0.0000 \\
\hline
3     & 1     & $-$     & 0.5   & 1     & 343074.6047 & 343074.7347 & -0.1300 \\
4     & 0     & $-$     & 0.5   & 1     & 343194.3039 & 343194.4339 & -0.1300 \\
5     & 1     & $-$     & 1.5   & 1     & 344196.3356 & 344196.4655 & -0.1299 \\
6     & 2     & $-$     & 1.5   & 1     & 344381.9550 & 344382.0849 & -0.1299 \\
\hline
7     & 2     & $+$      & 1.5   & 2     & 1030229.8178 & 1030230.9249 & -1.1071 \\
8     & 1     & $+$      & 1.5   & 2     & 1030363.5553 & 1030364.6624 & -1.1071 \\
9     & 2     & $+$      & 2.5   & 2     & 1032168.8370 & 1032169.9441 & -1.1071 \\
10    & 3     & $+$      & 2.5   & 2     & 1032341.1483 & 1032342.2554 & -1.1071 \\
\hline
11    & 3     & $-$     & 2.5   & 3     & 2060535.9577 & 2060540.0064 & -4.0487 \\
12    & 2     & $-$     & 2.5   & 3     & 2060675.3485 & 2060679.3973 & -4.0488 \\
13    & 3     & $-$     & 3.5   & 3     & 2063276.7730 & 2063280.8218 & -4.0488 \\
14    & 4     & $-$     & 3.5   & 3     & 2063443.5527 & 2063447.6015 & -4.0488 \\
\hline
15    & 4     & $+$     & 3.5   & 4     & 3433222.1380 & 3433231.9781 & -9.8401 \\
16    & 3     & $+$      & 3.5   & 4     & 3433364.6194 & 3433374.4596 & -9.8402 \\
17    & 4     & $+$      & 4.5   & 4     & 3436759.8067 & 3436769.6469 & -9.8402 \\
18    & 5     & $+$     & 4.5   & 4     & 3436923.5400 & 3436933.3802 & -9.8402 \\
\hline
19    & 5     & $-$     & 4.5   & 5     & 5147267.6517 & 5147285.8407 & -18.1890 \\
20    & 4     & $-$     & 4.5   & 5     & 5147412.0861 & 5147430.2751 & -18.1890 \\
21    & 5     & $-$     & 5.5   & 5     & 5151599.9592 & 5151618.1483 & -18.1891 \\
22    & 6     & $-$     & 5.5   & 5     & 5151761.7609 & 5151779.9499 & -18.1890 \\
\hline
23    & 6     & $+$      & 5.5   & 6     & 7201400.1636 & 7201426.5351 & -26.3715 \\
24    & 5     & $+$      & 5.5   & 6     & 7201545.9449 & 7201572.3164 & -26.3715 \\
25    & 6     & $+$      & 6.5   & 6     & 7206525.9256 & 7206552.2971 & -26.3715 \\
26    & 7     & $+$      & 6.5   & 6     & 7206686.3922 & 7206712.7637 & -26.3715 \\
\hline
27    & 7     & -     & 6.5   & 7     & 9594096.6941 & 9594124.3704 & -27.6763 \\
28    & 6     & -     & 6.5   & 7     & 9594243.4608 & 9594271.1371 & -27.6763 \\
29    & 7     & -     & 7.5   & 7     & 9600015.2023 & 9600042.8786 & -27.6763 \\
30    & 8     & -     & 7.5   & 7     & 9600174.6909 & 9600202.3672 & -27.6763 \\
\hline
31    & 8     & +     & 7.5   & 8     & 12323585.3054 & 12323594.8594 & -9.5540 \\
32    & 7     & +     & 7.5   & 8     & 12323732.8245 & 12323742.3785 & -9.5540 \\
33    & 8     & +     & 8.5   & 8     & 12330296.1028 & 12330305.6568 & -9.5540 \\
34    & 9     & +     & 8.5   & 8     & 12330454.8439 & 12330464.3979 & -9.5540 \\
\hline
35    & 9     & -     & 8.5   & 9     & 15387847.1770 & 15387798.6594 & 48.5176 \\
36    & 8     & -     & 8.5   & 9     & 15387995.2894 & 15387946.7718 & 48.5176 \\
37    & 9     & -     & 9.5   & 9     & 15395349.9512 & 15395301.4336 & 48.5176 \\
38    & 10    & -     & 9.5   & 9     & 15395508.1024 & 15395459.5848 & 48.5176 \\
\hline
\end{tabular}
\end{table}

\begin{table}
\caption{Comparison of \ce{^{24}Mg^1H} $\XSigma, v=0$ hyperfine line positions.
Five vibrational contracted basis functions $\ket{\XSigma,v=0, 1, 2, 3, 4}$ were
        used in this case.
    All frequencies are given in MHz.    }
    \label{tab:mghtransitionsv5}
\begin{tabular}{r|rrr|rrr|rrr}
\hline
No.   & $N'$     & $J'$     & $F'$     & $N''$     & $J''$     & $F''$     & $\nu_\duo$   & Measured (a) \cite{93ZiBaAn.MgH} & Measured (b) \cite{90ZiJeEv.MgH} \\
\hline
1     & 1     & 0.5   & 1     & 0     & 0.5   & 1     & 342997.636 & 342997.763(050) &  \\
2     & 1     & 0.5   & 0     & 0     & 0.5   & 1     & 343117.335 & 343117.463(050) &  \\
3     & 1     & 0.5   & 1     & 0     & 0.5   & 0     & 343305.510 & 343305.646(050) &  \\
4     & 1     & 1.5   & 1     & 0     & 0.5   & 1     & 344119.367 & 344119.497(050) &  \\
5     & 1     & 1.5   & 2     & 0     & 0.5   & 1     & 344304.986 & 344305.125(050) & 344305.3(20) \\
6     & 1     & 1.5   & 1     & 0     & 0.5   & 0     & 344427.241 & 344427.362(050) &  \\
\hline
7     & 2     & 1.5   & 2     & 1     & 0.5   & 1     & 687155.213 &       & 687157.17(17) \\
8     & 2     & 1.5   & 1     & 1     & 0.5   & 0     & 687169.251 &       & 687171.00(17)  \\
9     & 2     & 2.5   & 3     & 1     & 1.5   & 2     & 687959.193 &       & 687959.54(19)  \\
10    & 2     & 2.5   & 2     & 1     & 1.5   & 1     & 687972.501 &       & 687972.66(17)  \\
\hline
11    & 3     & 2.5   & 3     & 2     & 2.5   & 3     & 1028194.809 &       & 1028202.5(10)  \\
12    & 3     & 2.5   & 2     & 2     & 2.5   & 2     & 1028506.511 &       & 1028514.2(10)  \\
13    & 3     & 3.5   & 4     & 2     & 2.5   & 3     & 1031102.404 &       & 1031104.29(21) \\
14    & 3     & 3.5   & 3     & 2     & 2.5   & 2     & 1031107.936 &       & 1031104.29(21) \\
\hline
15    & 4     & 3.5   & 4     & 3     & 3.5   & 4     & 1369778.585 &       & 1369797.0(10) \\
16    & 4     & 3.5   & 3     & 3     & 3.5   & 3     & 1370087.846 &       & 1370107.5(10) \\
17    & 4     & 3.5   & 4     & 3     & 2.5   & 3     & 1372686.180 &       & 1372700.06(98) \\
18    & 4     & 3.5   & 3     & 3     & 2.5   & 2     & 1372689.271 &       & 1372700.06(98) \\
19    & 4     & 4.5   & 5     & 3     & 3.5   & 4     & 1373479.987 &       & 1373485.81(55) \\
20    & 4     & 4.5   & 4     & 3     & 3.5   & 3     & 1373483.034 &       & 1373485.81(55) \\
\hline
21    & 6     & 5.5   & 6     & 5     & 4.5   & 5     & 2054132.512 &       & 2054170.48(71) \\
22    & 6     & 5.5   & 5     & 5     & 4.5   & 4     & 2054133.859 &       & 2054170.48(71) \\
23    & 6     & 6.5   & 7     & 5     & 5.5   & 6     & 2054924.631 &       & 2054944.05(82) \\
24    & 6     & 6.5   & 6     & 5     & 5.5   & 5     & 2054925.966 &       & 2054944.05(82) \\
\hline
\end{tabular}
\end{table}

Finally, 
we list two calculated  $S$ branch ($\Delta J =2$) transitions
in the second and fourth rows of Table\,\ref{tab:mghRScompare}. 
These hyperfine-induced transitions are much weaker than the  two  $R$  branch ($\Delta J =1$) transitions in the first and third rows.

\begin{table}[htb]
    \centering
    \caption{Comparison of the line positions and strengths
    in the $R$ and $S$ branches of \ce{^{24}Mg^1H}
    $\XSigma, v=0$ hyperfine transitions.
    Line positions are given in \si{\MHz}.
    Five vibrational contracted basis functions $\ket{\XSigma,v=0, 1, 2, 3, 4}$ were used in this case.
    The line strength, $S$ [\si{Debye^2}], has the same definition as that in PGOPHER
    when the intensity unit option of PGOPHER,
    \texttt{IntensityUnit}, is chosen 
    as \texttt{HonlLondon}
    and the transition dipole moment
    is set to \SI{1}{Debye}.}
    \label{tab:mghRScompare}
\begin{tabular}{r|rrr|rrr|rr|rr}
\hline
No. & $F'$     & $\parity'$ & $J'$    & $F''$     & $\parity''$ & $J''$    & $\nu_\duo$    & $\nu_\mathrm{PG}$ & $S_\duo$ & $S_\mathrm{PG}$ \\
\hline
1     & 2     & $+$    & 2.5   & 1     & $-$     & 1.5    & 687972.5015 & 687973.4786 & 1.7558441 & 1.7558510 \\
2     & 2     & $+$     & 2.5   & 1     & $-$     & 0.5   & 689094.2323 & 689095.2094 & 0.0053314 & 0.0053315 \\
\hline
3     & 3     & $-$    & 3.5   & 2     & $+$     & 2.5    & 1031107.9360 & 1031110.8777 & 2.8371019 & 2.8371270 \\
4     & 3     & $-$     & 3.5   & 2     & $+$     & 1.5     & 1033046.9552 & 1033049.8969 & 0.0014804 & 0.0014805 \\
\hline
\end{tabular}%

  \end{table}
  
\section{Conclusion}

We demonstrate an algorithm for the calculation of 
hyperfine structure of diatomic molecules based on a
variational treatment of nuclear motion.
Nuclear magnetic dipole coupling terms
including Fermi-contact,
nuclear spin-electron spin dipole-dipole interaction,
nuclear spin - orbit,
nuclear spin - rotation,
and nuclear electric quadrupole interaction terms
are considered in our calculation.
New modules for the hyperfine structure calculation 
are added to the flexible variational nuclear-motion package \duo \cite{jt609}{}.

Based on the eigenfunctions and eigenvalues of $\bmJ$,
a parity-conserved rovibronic Hamiltonian matrix
of particular total angular momentum, $\bmF$,
is constructed and diagonalized.
The hyperfine wavefunctions
are finally represented using a Hund's case (a$_\beta$)
basis set.
Hyperfine-resolved line lists for diatomic molecules can be
 computed depending on the hyperfine
energy levels and wavefunctions.
To test the new module,
we calculate the hyperfine structure of 
the $v=0$, $\XSigma $ state of \ce{^{24}MgH}.
The results of \duo and PGOPHER show
excellent agreement for both line positions and line strengths.
The \duo code and the input file used for \ce{^{14}N^{16}O} and \ce{^{24}MgH}
are available at 
\href{https://github.com/ExoMol/Duo}{https://github.com/ExoMol/Duo}\,.

Our newly developed methodology builds a bridge
between calculations of electronic motion and
nucleus motion of diatomic molecules
which makes it possible to 
calculate nuclear magnetic dipole and electric quadruple
hyperfine structure effects from first principles.
Some hyperfine coupling constants considered in this work
may be calculated by quantum chemistry programs
\eg, DALTON \cite{14DALTON.method} and CFOUR \cite{CFOUR.methods}{}.
It is also possible to evaluate them manually after
obtaining  electronic wavefucntions \cite{05FiMaWe.hyperfine}{}.
We will discuss the \abinitio calculation
of hyperfine coupling constants in future work.

The current implementation only allows for nuclear spin effects
on  one atom and neglects coupling between electronic states.
The hyperfine coupling between two
electronic states is known to be important for some molecules.
For instance, to analyze the spectrum of \ce{I^{35}Cl},
\citeauthor{95SlClJa.hyperfine} also included
the hyperfine coupling terms 
between $\mathrm{X}\,^1\Sigma^+$ and $\mathrm{A}\,^3\Pi$
states \cite{95SlClJa.hyperfine}{}. 
Implementing this effect in \duo 
would require some further work on the matrix elements
but should not be a major undertaking.
Treating the case where both atoms possess a nuclear spin
introduce another source of angular momentum and the  
interaction between the two nuclei
also introduces new matrix elements \cite{78BrViLe.hyperfine}{}.
Here there are two possibilities, homonuclear systems, such as $^1$H$_2$ or $^{14}$N$_2$,
can be treated by generalizing the scheme given in this paper.
Heteronuclear systems, such as $^1$H$^{14}$N, are a little more complicated as they
give rise to different possible coupling schemes \cite{93Kato.methods}{}.
Our plan is to gradually update \duo for each of these cases
as the need for arises.

\begin{acknowledgement}

Qianwei Qu acknowledges the financial support from
University College London and China Scholarship Council.
This work was supported by the STFC Projects 
No. ST/M001334/1 and ST/R000476/1, 
and ERC Advanced Investigator Project 883830.

\end{acknowledgement}

\begin{suppinfo}

\duo is an open-source software, 
which is available at 
\href{https://github.com/ExoMol/Duo}{https://github.com/ExoMol/Duo}\,, where
\duo input files used in this work can be found.
The \duo and PGOPHER input files used in this work are also provided as supplementary data.

\end{suppinfo}

\bibliography{bib_journals_iso,bib_hyperfine,bib_jtj,bib_methods,bib_NO,bib_MgH,bib_VO,bib_sy.bib}

\end{document}